\documentclass[aps,prx,reprint,superscriptaddress, longbibiography]{revtex4-1}
\usepackage{color}
\usepackage{soul}
\usepackage{dcolumn}
\usepackage{graphicx}
\usepackage{amsmath}
\usepackage{amssymb}
\usepackage{bm}

\begin{document}




\title{Protecting a superconducting qubit from energy decay by selection rule engineering}




\author{Yen-Hsiang Lin, Long B. Nguyen, Nicholas Grabon, Jonathan San Miguel, Natalya Pankratova, and Vladimir E. Manucharyan}
\affiliation{Department of Physics, Joint Quantum Institute, and Center for Nanophysics and Advanced Materials,
University of Maryland, College Park, MD 20742, USA.}



\date{\today}

\begin{abstract}
Quantum control of atomic systems is largely enabled by the rich structure of selection rules in the spectra of most real atoms. Their macroscopic superconducting counterparts have been lacking this feature, being limited to a single transition type with a large dipole. Here we report a superconducting artificial atom with tunable transition dipoles, designed such that its forbidden (qubit) transition can dispersively interact with microwave photons due to the virtual excitations of allowed transitions. Owing to this effect, we have demonstrated an in-situ tuning of qubit's energy decay lifetime by over two orders of magnitude, exceeding a value of $2~\textrm{ms}$, while keeping the transition frequency fixed around $3.5~\textrm{GHz}$
\end{abstract}


\maketitle



Out of the many quantum information processing platforms~\cite{ladd2010quantum}, circuit quantum electrodynamics (QED) stands out because macroscopic superconducting artificial atoms have a natural tendency to interact strongly with radiation~\cite{wallraff2004strong,schoelkopf2008wiring}. As a price for strong interactions, a circuit atom suffers from energy decay by spontaneous emission into its diverse solid-state dissipative environment~\cite{leggett1980macroscopic}. To give a few examples, a microwave photon emitted by a superconducting qubit can be absorbed by out-of-equilibrium quasiparticles in superconducting leads (quasiparticle loss)~\cite{Martinis2009quasiparticle, Catelani2011quasiparticle2, wang2014measurement, pop2014coherent}; by two-level defects in the oxide layers (dielectric loss)~\cite{martinis2005decoherence,wang2015surface}; or simply by spurious electromagnetic modes of the measurement circuitry (Purcell effect)~\cite{houck2008controlling}. Controlling such a complex environment is a tremendous task, and, despite impressive progress~\cite{paik2011observation, barends2013coherent}, energy decay remains a major limitation for circuit QED systems~\cite{devoret2013superconducting}. Therefore, it is tempting to find a way to inhibit energy decay of a qubit by decoupling it from an arbitrary environment without losing quantum control over it.

An interesting example of such a system is an optical atomic clock~\cite{ludlow2015optical}. Its host atom (or ion) has a highly controlled and long-lived clocking transition despite the fact that the atom faces a fixed dissipative environment -- the vacuum. This is made possible by a rich structure of selection rules in the spectra of many even hydrogen-like atoms and ions. For instance, the transition $S_{1/2} - P_{1/2}$ of a Ca$^+$ ion has a relatively large transition dipole and belongs to the ``allowed" type with a radiative lifetime of the order of $10~\mathrm{ns}$. A similar frequency  transition (visible) $S_{1/2} - D_{5/2}$ sees the same radiative environment, and yet its lifetime exceeds $1~\mathrm{s}$. This transition is ``forbidden" as its dipole is nulled by the symmetry of atomic wave functions, and the long lifetime is due to a much weaker quadrupolar coupling. One can then encode a qubit onto the forbidden (clocking) transition and yet efficiently read it out by measuring the state-dependent fluorescence of the allowed transition. Other allowed transitions are often utilized for qubit state preparation by optical pumping and two-tone Raman drives~\cite{cirac1995quantum}. Moreover, even a two-qubit remote gate operation can be performed on a pair of decoupled atomic clock qubits by a joint quantum measurement~\cite{moehring2007entanglement}. 

Inspired by this atomic clock example, one may ask the following questions with regard to superconducting qubits: (i) can we engineer a forbidden transition in a superconducting circuit, and how would its quality factor $Q_1$ grow upon reducing the transition dipole? (ii) can a forbidden transition still interact with a cavity mode in order to perform measurement and multi-qubit operations? Unfortunately, common superconducting circuits lack the selection rule diversity of real atoms. In fact, their spectra correspond to either a two-level system or a weakly anharmonic oscillator, and thus are limited to only one transition type~\cite{clarke2008superconducting}. 
 
In this work we have designed a fluxonium artificial atom~\cite{Manucharyan113} such that it combines both allowed and forbidden transitions. Moreover, unlike real atoms, here the transition dipoles are continuously and broadly tunable by magnetic flux. We show that even with a vanishing transition dipole, a qubit still undergoes a finite and purely dispersive (longitudinal) interaction with cavity photons, enabled by the other atomic transitions with large dipoles. This effect allowed us to explore energy relaxation in our circuit in the regime of extreme decoupling from environment. The measured quality factor of a qubit transition, deliberately confined to a narrow range near $3.5~\textrm{GHz}$, scales linearly with the inverse square of the transition's dipole, and reaches a benchmark value of $Q_1 > 4\times 10^7$ corresponding to $T_1 > 2~\textrm{ms}$.

	\begin{figure*}
		\centering
		\includegraphics[width = \linewidth]{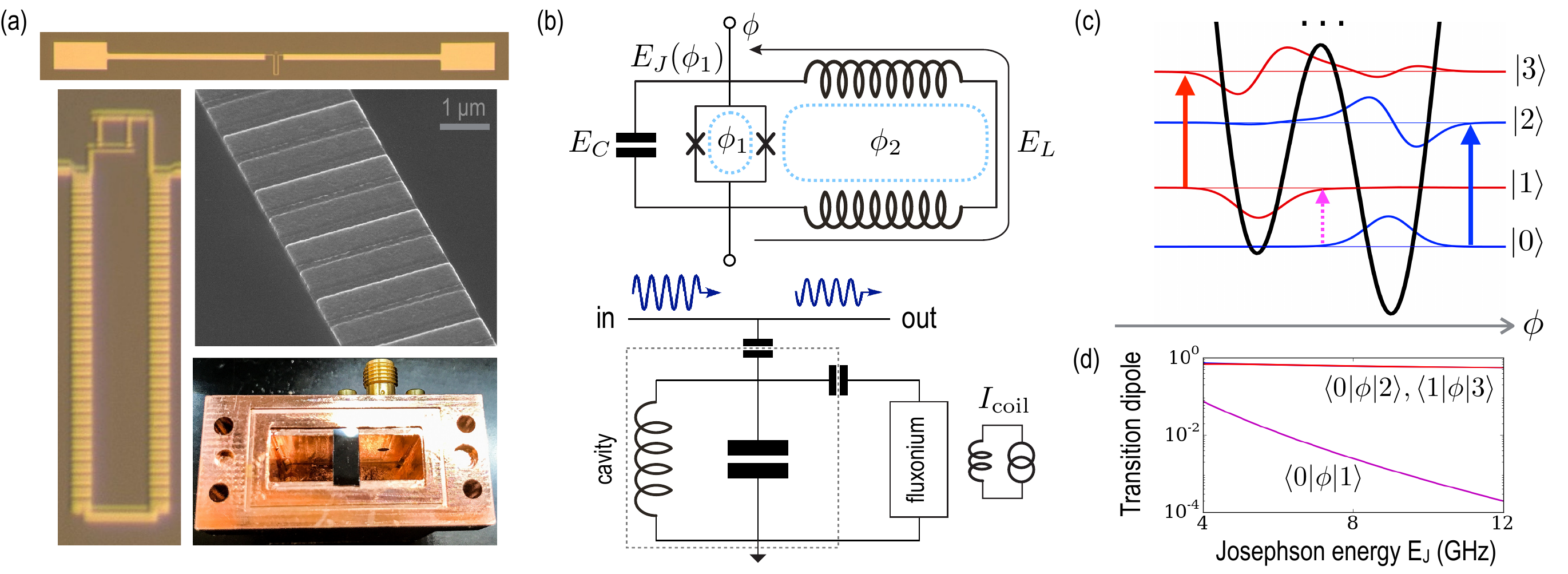}
		\caption{(\textbf{a}) Images of a double loop fluxonium device in a 3D cavity. The SQUID loop is much smaller in area than the main loop, and the antenna is connected directly to the split-junction.  (\textbf{b}) Circuit models of a double-loop fluxonium atom (top) and its coupling to a 3D copper cavity (bottom). (\textbf{c}) Lowest 4 energy levels of the atom accurately positioned in the double-well potential profile, $U(\phi)$, along with their calculated wave-functions.
			  (\textbf{d}) The inter-well \textit{fluxon} transition $0-1$ (magenta arrow in (c)) has a vanishing transition dipole $\langle 0|\phi|1\rangle$ and hence is of the ``forbidden" type. The intra-well \textit{plasmon} transitions $0-2$ and $1-3$ (blue and red arrows in (c), respectively) by contrast have transition dipoles of order unity and are thus of the ``allowed" type.}
		
		\label{fig:Fig1}
	\end{figure*}

Compared to original fluxonium design, here we have replaced the single weak junction with a flux-tunable split-junction and coupled it capacitively to a 3D copper cavity using a $\textrm{mm}$-size antenna (Fig.~\ref{fig:Fig1}a). The resulting modified fluxonium can be viewed as a 3D transmon shunted by a large linear inductance of a Josephson array (Fig.~\ref{fig:Fig1}b). The circuit Hamiltonian, given by Eq.~(\ref{Eq: Hamiltonian}), is defined by the inductive energy of the Josephson chain $E_L$, the charging energy of the total capacitance $E_C$, and the variable Josephson energy of the split-junction $E_J(\phi_1)$, as well as by the two fluxes $\phi_1$ and $\phi_2$ piercing the split-junction and the main loops, respectively. The generalized flux $\phi$ in the inductance is a position-like quantum variable of the circuit (we take all fluxes in units of $\hbar/2e$). It moves in an effective potential given by $U(\phi) = E_L\phi^2/2 - E_J(\phi_1)\cos(\phi+\phi_J(\phi_1)-\phi_2)$, where $E_J(\phi_1)$ and $\phi_J(\phi_1)$ are given by Eq.~(\ref{Eq:SQUID}). Kinetic energy is given by the term $4E_C n^2$, where $n = -i\partial_{\phi}$ is a momentum-like continuous variable conjugate to $\phi$.

Two distinct transition types emerge in our circuit in the regime $E_L/E_J \ll 1$ and $E_J/E_C \gtrsim 10$. The former condition ensures that the potential $U(\phi)$ consists of multiple Josephson wells, whose depth and elevation are tuned by the external fluxes $\phi_1$ and $\phi_2$, respectively. The latter condition weakens quantum tunneling such that every low energy state of the circuit tends to localize inside a single well (Fig.~\ref{fig:Fig1}c). The intra-well transitions are called \textit{plasmons} by analogy with plasma oscillations in junctions. Similar transitions occur in a transmon~\cite{koch2007charge}, except that here a plasmon remains charge-insensitive even for a small value of $E_J/E_C$ due to the inductive shunt~\cite{koch2009charging}. The inter-well transitions are called \textit{fluxons}. These are accompanied by a twist in the superconducting phase along the fluxonium main loop by $2\pi$. A fluxon is analogous to the transition of a flux qubit~\cite{Paauw2009Tuning}, except that it is about $10^2 -10^3$ times less sensitive to flux noise due to the large number of junctions in the fluxonium loop~\cite{manucharyan2012evidence}. As long as the two adjacent wells are offset against each other, the two states connected by a fluxon would have a vanishing overlap~\cite{kerman2010metastable}. Fluxon is therefore a ``strongly forbidden" transition in the sense that any operator $O(\phi)$ would have an exponentially small matrix element for sufficiently large ratio $E_J/E_C$. By contrast, plasmons are ``strongly allowed" (Fig.~\ref{fig:Fig1}d), because their transition dipoles, naturally defined as matrix elements of $\phi$, are all near unity for a broad range of values of $E_J/E_C$~\cite{zhu2013asymptotic}.

Because of the vanishing transition dipole, the transverse interaction of a fluxon with a cavity mode is negligible. However, we found that there is a purely dispersive longitudinal interaction between the two in the form of $H_{\textrm{int}} = \chi \sigma_z a^{\dagger} a$, where $a$ is the photon annihilation operator, and $\sigma_z$ is the fluxonium Hamiltonian projected onto its two eigenstates connected by a fluxon~(see appendix~\ref{appendix:cQED}). The origin of a finite dispersive shift $\chi$ can be understood as follows. In the state $0$, which can be approximately viewed as the vibrational ground state of the lower well, fluxonium shifts the bare cavity resonance by an amount $\chi_0$, due to virtual excitations of the lower well (blue) plasmon. The shift $\chi_0$ grows as the plasmon frequency approaches the cavity resonance, and has a relatively large magnitude similar to that of a typical  transmon qubit. Analogously, the state $1$ shifts the cavity by an amount $\chi_1$, due the higher well (red) plasmon. Since the lower and higher well plasmons have different frequencies, $\chi_0 \neq \chi_1$, giving rise to a finite dispersive shift $\chi = \chi_1 - \chi_0$. Quantitatively, the values of $\chi_{0(1)}$ are found by summing contributions from virtual excitations of every transition starting from the states $0(1)$, and in general there is no reason for the two to be equal~\cite{zhu2013circuit}. 


	\begin{figure*}
		\centering
		\includegraphics[width = \linewidth]{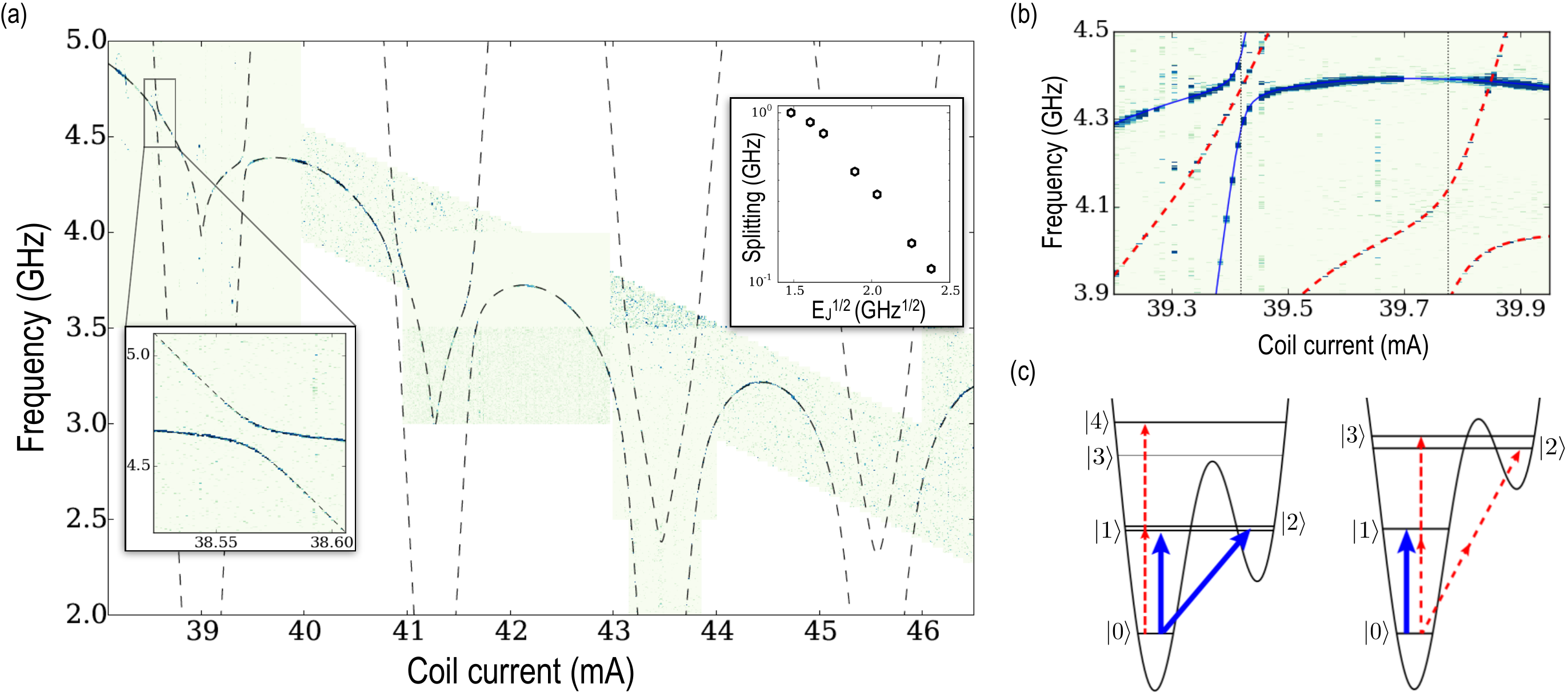}
		\caption{\textbf{(a)} Transmission near cavity resonance (scale not shown) as a function of the coil current and spectroscopy tone frequency. Cavity resonance is off the scale at about $10~\mathrm{GHz}$. Dashed lines represent a fit to the circuit model in Fig.~\ref{fig:Fig1}b (see text). Left inset: a zoom-in on the smallest fluxon-plasmon splitting region. Right inset: values of seven splittings visible in the main plot vs. calculated value of $E_J^{1/2}$. \textbf{(b)} A section of spectroscopy data from (a) retaken at a higher spectroscopy tone power, which reveals two-photon processes. Theory prediction for the two-photon resonances matches data without fitting (dashed red lines). \textbf{(c)} Illustrations of the one-photon and two-photon splittings visible in (b), created with extracted circuit parameters.}
		
		\label{fig:Fig2}
	\end{figure*}

The spectrum of our artificial atom is revealed by a two-tone transmission spectroscopy signal measured as a function of the spectroscopy tone frequency and the current in the external coil that creates a global flux bias (Fig.~\ref{fig:Fig2}). The readout tone was irradiated near the cavity's resonance at $10~\mathrm{GHz}$. Due to linearity of the coil, it is safe to assume that $\phi_{1,2} = \beta_{1,2} I_{\textrm{coil}}$, where $I_{\textrm{coil}}$ is the coil current and $\beta_{1,2}$ are flux couping constants. The two observed resonances vary with the coil current in a sophisticated quasi-periodic manner (Fig.~\ref{fig:Fig2}a). Nevertheless, the two fit remarkably well to the numerically-obtained lowest transitions of the circuit Hamiltonian (Eq.~\ref{Eq: Hamiltonian}) involving only six adjustable parameters: $E_C, E_L, E_{J_1}, E_{J_2}, \beta_1, \beta_2$. Here $E_{J_{1,2}}$ are the Josephson energies of the two junctions in the SQUID. Having established a tight correspondence between the data and a simple circuit model of our atom, we now proceed to interpret the most essential spectral features of the data in Fig.~\ref{fig:Fig2}a.

The quasi-periodicity of the spectrum as a function of the coil current corresponds to changing the external flux in the main loop by a flux quantum, i.e. $\phi_2 \rightarrow \phi_2 + 2\pi$. The SQUID loop has a much smaller area and hence a much larger period. The point of inversion symmetry of the spectrum at $I_{\textrm{coil}} \approx 45.5~\mathrm{mA}$ corresponds to biasing the SQUID loop with a half a flux quantum, i.e $\phi_1= \pi$, and the Josephson energy $E_J(\phi_1=\pi) = |E_{J_1}-E_{J_2}|$ reaches its minimum. The separation of the spectrum into fluxons and plasmons is particularly apparent in the region $38~\mathrm{mA}<I_{\textrm{coil}}<42~\mathrm{mA}$. The weakly flux-dependent transition with multiple sweet-spots is the lower-well plasmon. Due to the presence of the inductive shunt, plasmon's frequency is not a monotonic function of $E_J$, although it reduces with $E_J$ on average. The transition that changes linearly with the coil current in a zigzag pattern is a fluxon. The avoided crossings correspond to a full hybridization of a fluxon with a plasmon (Fig.~\ref{fig:Fig2}a, left inset). The frequency splitting quantifies the strength of inter-well transitions, varying from $100~\mathrm{MHz}$ at $I_{\textrm{coil}}=38.56~\mathrm{mA}$, where a fluxon is well defined, to over $1~\mathrm{GHz}$ near $\phi_1 = \pi$, where this notion becomes vague. The top inset of Fig.~\ref{fig:Fig2}a illustrates that at sufficiently large values of $E_J$, the logarithm of the splitting scales as $E_J^{1/2}$, in agreement with the WKB description of tunneling. 

	\begin{figure}
		\centering
		\includegraphics[width=\linewidth]{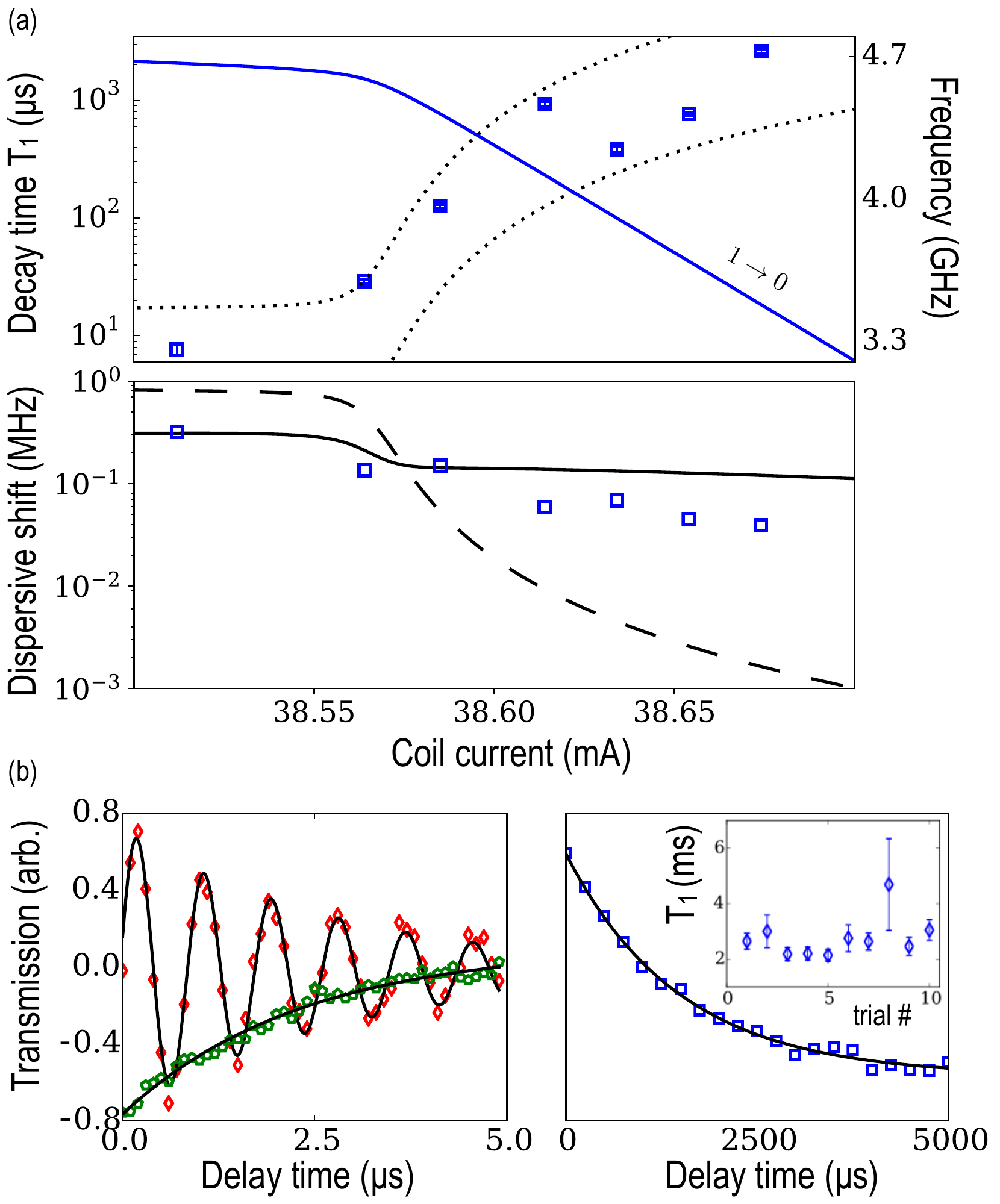}
		\caption{(\textbf{a}) The measured lifetime $T_1$ of the $0-1$ transition (markers) and its frequency (solid line). Dotted lines are the dielectric loss prediction corresponding to the loss tangent values of $2\times10^{-5}$ and $2\times10^{-6}$. Measured dispersive shift (blue markers) and calculated dispersive shifts, taking into account all transitions to higher levels (solid line) and only states $0$ and $1$ (dashed line). Note that $\chi$ decreases much slower than $T_1$ grows. (\textbf{b}) Example of time-domain measurements: Rabi oscillation trace (red), $\pi$-pulse echo trace (green), and energy relaxation trace (blue) all measured simultaneously with an interleaved pulse sequence. The inset shows periodic measurements of $T_1$ during a period of about one hour.
			}
		\label{fig:Fig3}
	\end{figure}

Two-photon resonances appear at a higher spectroscopy tone power and match theory without adjustable parameters (Fig.~\ref{fig:Fig2}b). Thus, at $I_{\textrm{coil}} \approx 39.4~\mathrm{mA}$, we observe a two-photon transition to the 4th excited state of the circuit, which lies above the barrier. Coincidentally, this two-photon resonance landed right inside the small frequency gap formed by the conventional one-photon fluxon-plasmon splitting. As the offset between the two wells increases further, we observe a two-photon avoided crossing of the second and third excited states at $I_{\textrm{coil}} \approx 39.8~\mathrm{mA}$. It corresponds to a hybridization of a fluxon with a two-plasmon excited state of the lower well. The two-photon data also reveals that the plasmon's anharmonicity is about $700~\textrm{MHz}$, a value significantly larger than that of conventional transmons, but similar to recently developed capacitively-shunted flux qubits~\cite{yan2016flux}.


In order to proceed to analysis of energy relaxation data, we also need the knowledge of transition dipoles of our artificial atom. Measuring them accurately is hard. Instead, we can simply calculate them using the parameters of the circuit Hamiltonian obtained from fitting the spectroscopy data. This approach is justified by the fact that our simple four-parameter Hamiltonian (Eq.~\ref{Eq: Hamiltonian}) appended by the two current-to-flux conversion constants $\beta_{1,2}$, matches multiple atomic transitions with intricate current dependence over a large range frequencies and has a power-broadened line-width not exceeding $0.1\%$.

Controlled inhibition of energy decay in our circuit is most clearly demonstrated by measuring the lifetime of the lowest $0-1$ transition as we tune the coil current through the plasmon-fluxon anticrossing, shown in the lower inset of Fig.~\ref{fig:Fig2}a. Indeed, we observed a drastic enhancement of the $T_1$ time from $T_1 < 10~\mathrm{\mu s}$ at the plasmon side to $T_1 > 1~\mathrm{ms}$ at the fluxon side (Fig.~\ref{fig:Fig3}a). To interpret the data quantitatively, we turned to the model of dielectric loss, commonly encountered in transmon qubits~\cite{wang2015surface}. This model echoes the observed enhancement of $T_1$ and requires the bounds on the effective loss tangent of the split-junction to lie between $2\times 10^{-5}<\tan\alpha < 2\times 10^{-6}$. For the same range of coil currents, we plot the calculated dispersive shift $\chi$, which remarkably does not drop significantly at the fluxon side of the anticrossing (Fig.~\ref{fig:Fig3}b). The measured dispersive shift, extracted from the Rabi oscillations amplitude, agrees reasonably well with theory (Eq.~\ref{Eq: dispersive shift}), without using any assumptions. In sharp contrast, the dispersive shift, calculated taking into account only the  $0$ and $1$, drops rapidly with the increase of $T_1$, thereby emphasizing the importance of plasmons in enabling the dispersive circuit QED with a forbidden transition.

Rabi oscillations along with $\pi$-pulse echo experiments demonstrate that a fluxon remains coherent even when its transition dipole suppressed to the extent that $T_1 > 1.5~\textrm{ms}$ (Fig.~\ref{fig:Fig3}c). The coherence time $T_2$, given by the characteristic decay time of the echo signal, is given by $T_2 \approx 2~\mu s$, and is likely limited by the first-order flux noise in both the main and the SQUID loops. The decay of the readout signal following a $\pi$ pulse on a qubit fits the exponential function well and repeated experiments did not produce more than a factor of two variation of $T_1$ within about one hour (Fig.~\ref{fig:Fig3}c - inset). This leads us to believe that the fluctuation of $T_1$ values in Fig.~\ref{fig:Fig3}a occurs on longer than a one hour time scale.

	\begin{figure}
		\centering
		\includegraphics[width=\linewidth]{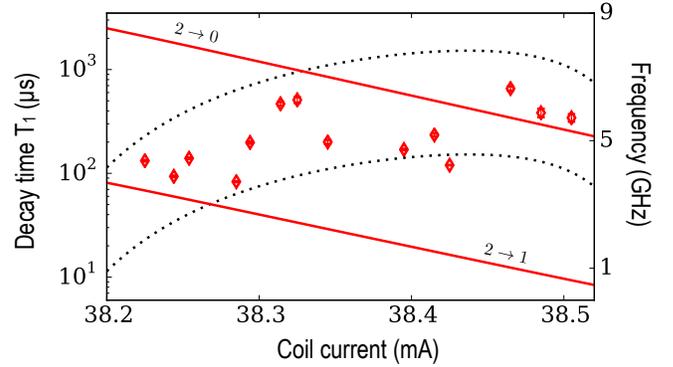}
		\caption{Measured energy decay time of the second excited state (markers) and the $0-2$ and $1-2$ transition frequency (solid lines). The transition $1-2$ dominates the decay in the given interval of coil currents. Dashed lines represent the dielectric loss for the same parameters as in Fig.~\ref{fig:Fig3}a }
		
		\label{fig:Fig4}
	\end{figure}

For a control experiment, we note that the transitions $1-2$ and $0-2$ become forbidden at the plasmon side of the anticrossing, i.e. for $I_{\textrm{coil}} < 38.55~\textrm{mA}$. To test this idea, we have measured the decay of the readout signal following a $\pi$-pulse applied to the $0-2$ transition (Fig.~\ref{fig:Fig4}). Again, we observed long relaxation times reaching values over $600~\mathrm{\mu s}$ at a point corresponding to the $0-2$ transition frequency of about $6.5~\textrm{GHz}$ and $2-1$ transition frequency of about $2.8~\textrm{GHz}$. The measured $T_1$ values, as a function of $I_{\textrm{coil}}$, fall within the bounds imposed by the same dielectric loss models previously applied to the $0-1$ decay data at the other side of the anticrossing (Fig.~\ref{fig:Fig3}). We also note, that the direct decay channel $2-0$ turns out to be negligible compared to the indirect channel $2-1-0$. Given that the transition $1-0$ here decays much faster than $2-1$, we therefore associate the decay of the $2$-state with the environment at the frequency of the $2-1$ transition.

\begin{figure}
	\centering
	\includegraphics[width=\linewidth]{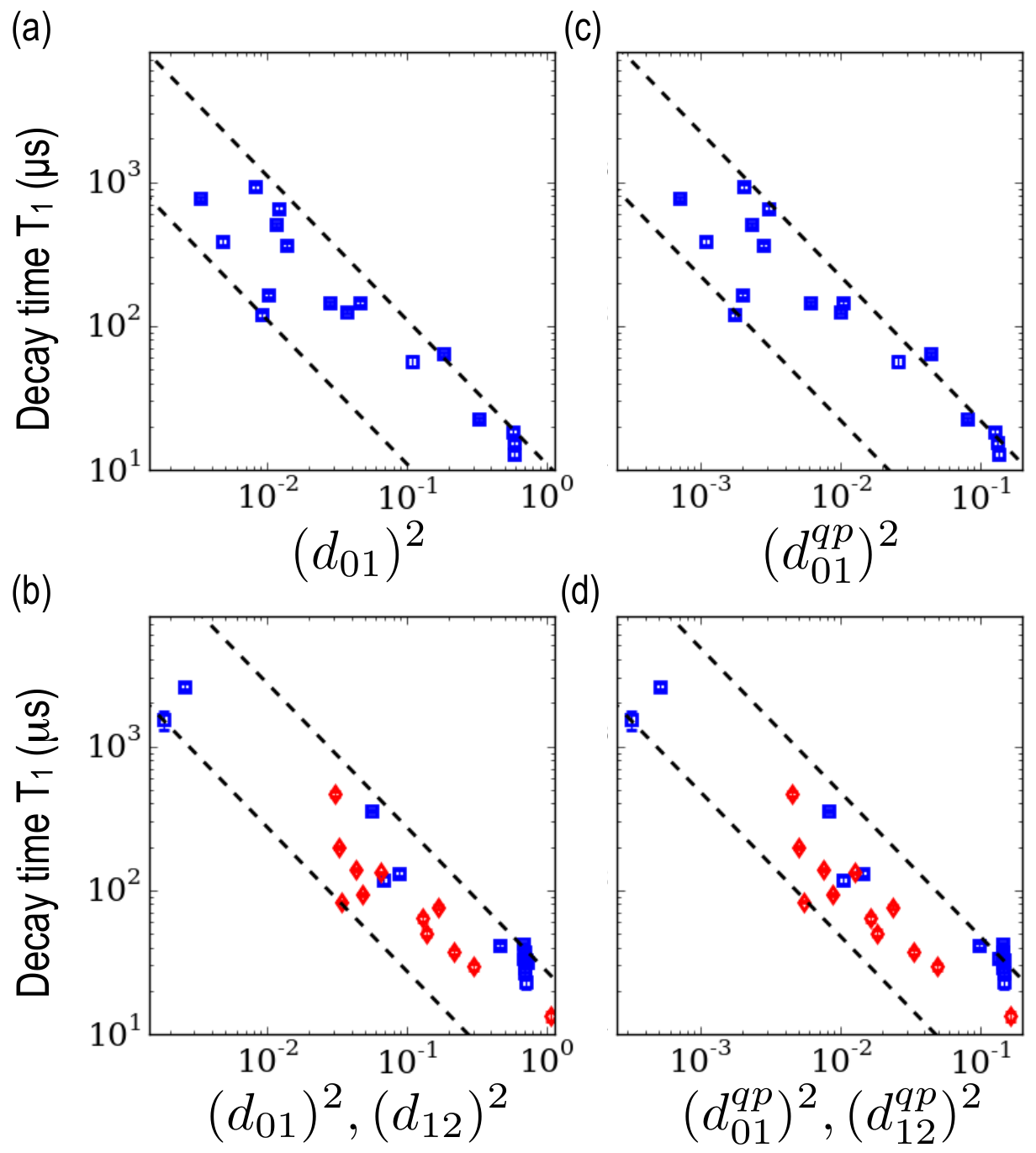}
	\caption{
		Energy decay times for transitions $0-1$ (blue) and $1-2$ (red) taken from a narrow range of frequencies vs. the corresponding transition dipole squared. (\textbf{a,c}) Frequency range $3.5-4.5~\textrm{GHz}$, dashed lines represent loss tangent bounds of $2\times 10^{-6} <\tan\alpha <2\times 10^{-5}$ for (a) and normalized quasiparticle density bounds of $5\times 10^{-7}< x_{qp} < 5\times 10^{-6}$ for (c). (\textbf{b,d}) Frequency range $2.5-3.5~\textrm{GHz}$, dashed lines represent loss tangent bounds of $1.5\times 10^{-6} <\tan\alpha <1.5\times 10^{-5}$ for (b) and normalized quasiparticle density bounds of $2\times 10^{-7}< x_{qp} < 2\times 10^{-6}$ for (c). The data in (\textbf{c,d}) is plotted vs. the effective transition dipole for non-linear coupling to quasiparticle tunneling in the split-junction, defined in Eq.~(\ref{Eq: DipoleQP}).
	}
	
	\label{fig:Fig5}
\end{figure}

Finally, we have collected energy decay times for the lowest transition taken at a number of special values of $I_{\textrm{coil}}$ such that the transition dipole $d_{01}$, given by $d_{01} = \langle 0|\phi|1\rangle$, vastly varies while the transition frequency is confined to a narrow interval $3.5-4.5~\textrm{GHz}$. All such transitions lose energy to the essentially identical environment. Therefore, Fermi's golden rule predicts that at zero temperature $1/T_1^{1\rightarrow0} \propto (d_{01})^2$ for an arbitrary linearly-coupled environment. Our data obeys this simple scaling for the values of $T_1$ spanning a remarkable range of over two orders of magnitude (Fig.~\ref{fig:Fig5}a). Despite some fluctuations, the data clearly shows that the dramatic enhancement of $T_1$ of a qubit occurs solely due to the reduction of its transition dipole. Because the suppression of $d_{01}$ deep in the fluxon regime has no classical analog, the observed scaling evidences that the energy decay occurs by a spontaneous emission rather than by a thermal activation. Data for the neighboring $2.5 - 3.5~\textrm{GHz}$ frequency range, including relaxation of the $2-1$ transition, confirms our conclusion (Fig.~\ref{fig:Fig5}b).

The only known non-linear loss mechanism in which coupling to the bath cannot be described using the matrix elements of $\phi$ is quasiparticle tunneling across the small junctions~\cite{Catelani2011quasiparticle1,pop2014coherent}. The effective transition dipole $d_{ij}^{qp}$ is responsible for the coupling to quasiparticles turns out to be a more complex function of $\phi$, given by Eq.~(\ref{Eq: DipoleQP}). Plotting the measured $T_1$'s as a function of the calculated value of $(d^{qp}_{ij})^2$, we find the same as for the linear dissipation (Fig.~\ref{fig:Fig5}c,d). The similarity of the two data sets, (a,b) and (c,d) of  Fig.~\ref{fig:Fig5}, reflects the simple fact that the fluxon transition is forbidden irrespective of the loss mechanism because of the vanishing overlap of the qubit states wave functions illustrated in Fig.~\ref{fig:Fig1}c

In conclusion, we have realized an artificial circuit atom with tunable selection rules and demonstrated a dispersive circuit QED with a strongly forbidden atomic transition. The observation of a quality factor as high as $Q_1 > 4\times 10^7$, without showing signs of saturation upon reducing the transition dipole, validates the general idea of protecting qubits from an arbitrary dissipative environment. Our specific method of inhibiting energy decay came with a disadvantage that the fluxon is first-order flux sensitive. Nevertheless, its coherence time was over $2~\mu s$ thanks to the large inductance of the Josephson array comprising the fluxonium main loop. Our experiment is thus a natural starting point for implementing more sophisticated protection schemes~\cite{ioffe2002possible}, e.g. those based around $0-\pi$ qubits, which rely on a similar mechanism for protecting against energy decay, but offer additional protection from dephasing and gate errors~\cite{brooks2013protected,bell2014protected}. 

Future improvements to this experiment include enhancing the fluxon's dispersive shift by arranging the plasmons closer to the cavity, utilizing Raman transitions to speed-up single-qubit gate operations, and increasing the chain inductance to further suppress flux noise. The combination of forbidden and allowed transitions in a single artificial circuit atom makes it useful for a range of quantum applications, from testing the limits of energy relaxation in macroscopic circuits and exploring the driven dynamics of meta-stable quantum systems to controlling photonic states of cavities~\cite{mirrahimi2014dynamically,vlastakis2013deterministically} and creating coherently interacting spin $1/2$ registers for quantum annealers~\cite{boixo2014evidence,weber2017coherent}.

\begin{acknowledgments}
	We would like to thank Nathana\"el Cottet for fruitful discussions. We acknowledge funding from US National Science Foundation (DMR-1455261) and ARO-MURI "Exotic states of light in superconducting circuits".
\end{acknowledgments}

\appendix
\section{Theoretical methods}

\subsection{Tunable Fluxonium Hamiltonian}
\label{A1}
The Hamiltonian of the fluxonium atom with a split-junction is given by
\begin{equation}
\label{Eq: Hamiltonian}
H_{\textrm{a}} = 4 E_C n^2 + E_L\frac{ \phi^2}{2} - E_J(\phi_1)\cos(\phi- \phi_J(\phi_1)-\phi_2).
\end{equation}
The charging energy $E_C = e^2/2C$ contains the total capacitance $C$ including the dominant antenna contribution that shunts the split-junction of the fluxonium. The inductive energy scale is defined as  $E_L = E_{JA}/N$, where $E_{JA}$ is the Josephson energy of the array junctions and $N$ is the total number.  The tunable Josephson energy $E_J(\phi_1)$ is given by the relations 
\begin{align}
\label{Eq:SQUID}
& E_J(\phi_1) = E_{J\Sigma}|\cos\frac{\phi_1}{2}|\sqrt{1+d^2 \tan^2\frac{\phi_1}{2}}\\
&\phi_J(\phi_1) =\arctan \large{(} d \tan\frac{\phi_1}{2}\large)\\
& E_{J\Sigma} = E_{J1} + E_{J2}, d = |E_{J1}-E_{J2}|/E_{J\Sigma}.
\end{align} 

\subsection{Circuit QED with a forbidden transition}\label{appendix:cQED}

The coupling of the atom to a 3D cavity mode is given by a transverse linear coupling in the form of 

\begin{equation}
\label{cQED}
H_{cQED} = H_{a} + \hbar \omega_c a^{\dagger}a + g n (a+a^{\dagger}),
\end{equation}
where $a$ is the photon annihilation operator, $\omega_c$ is the cavity mode frequency, and $g$ is an empirical coupling constant. 

Applying second-order perturbation theory, one obtains the following expression for the dispersive shift, $\chi_{j}$, of the cavity by the atomic state $j$:
\begin{equation}\label{Eq: dispersive shift}
\chi_{j} = \frac{g^2}{16E_C^2} \sum_{k\neq j} \frac{\omega_{jk}^3}{\omega_{jk}^2 - \omega_c^2} \times (d_{jk})^2.
\end{equation}
The transition dipoles $d_{jk}$ are given by $d_{jk} =\langle j|\phi|k\rangle$. Here we have taken into account that $\phi$ and $n$ are linearly related via $2e n = C(\hbar/2e)\dot{\phi}$.

The most important property of Eq.~(\ref{Eq: dispersive shift}) can be illustrated with Fig.~\ref{fig:Fig1}c. Namely, the fluxon's dispersive shift $\chi_{01} \neq 0$ even though $d_{01} \rightarrow 0$. Moreover, $\chi_{01}$ can be of the same order of magnitude as $\chi_{02,13}$ (plasmon shifts) if $\omega_{02}\neq\omega_{13}$ and either $\omega_{13} \approx \omega_c$ or $\omega_{02} \approx \omega_c$. This leads to a purely dispersive coupling of the fluxon transition $0-1$ to the cavity, given by an effective Hamiltonian

\begin{equation}
	H_{\textrm{fluxon-photon}} = \hbar(\omega_c +\chi_{01}\sigma_z)a^{\dagger}a.
\end{equation}
\subsection{Energy relaxation rates}
\subsubsection{Dielectric loss and other linear dissipation}
The linear coupling of an atomic transition to a specific environment is given by the interaction Hamiltonian $H_{int} = (\hbar/2e)\phi I$, where $I$ represents the collective environmental degree of freedom and $\phi$ is the generalized coordinate of the atom. According to Fermi's golden rule, relaxation rate is given by
\begin{equation}\label{Eq: GoldenRule}
1/T_1^{j\rightarrow i} = \frac{1}{\hbar^2}|d_{ij}|^2 S(\omega_{ij}),
\end{equation}
where $S(\omega_{ij})$ is the spectral density of noise of the environmental variable $I$ at the transition frequency $\omega_{ij}$. In the case of the commonly encountered dielectric loss~\cite{wang2015surface}, the dissipation presumably originates from filling the shunting capacitance $C$ with a lossy dielectric, characterized by a loss tangent, $\tan\delta$.  In that case $S(\omega) = 2(\hbar \omega)^2 R_Q C \tan\delta$, and
\begin{equation}\label{Eq: T1dielectric}
1/T_1^{j\rightarrow i}|_{\textrm{dielectric loss}} = 2\tan\delta R_QC\omega_{ij}^2 \times d_{ij}^2, 
\end{equation}
where we have introduced reduced superconducting resistance quantum, $R_Q = \hbar/4e^2 \approx 1~k\Omega$. 
Note that when the coupling is via the charge variable $n$ or a different relaxation mechanism (e.g. Purcell effect), the expression (\ref{Eq: GoldenRule}) still holds, but $S(\omega)$ will be a different function of frequency. 
\subsubsection{Quasiparticle loss}

Quasiparticle tunneling across the small junction of fluxonium has been recently identified as a peculiar non-linear dissipation source. Here the coupling Hamiltonian must be written as $H_{int} = (\hbar/2e)\sin\frac{\phi-\phi_2}{2}\times I_R + (\hbar/2e)\sin\frac{\phi-\phi_2-\phi_1}{2}\times I_L$ which includes independent tunneling processes in both the left and the right junctions of the SQUID. Where operators $I_{R,L}$ represent the environment seen by the two junctions. Assuming that the noise spectral densities of $I_L$ and $I_R$ are identical, the quasiparticle relaxation rate is given by 
\begin{equation}
1/T_1^{j\rightarrow i}|_{\textrm{quasiparticle tunneling}} = x^{qp}F(\omega)\times(d_{ij}^{qp})^2,
\end{equation}
where $F(\omega)$ is given in Ref.~\cite{Catelani2011quasiparticle2}, $x_{qp}$ is the fraction of broken Cooper pairs in the superconductor, and the effective transition dipole is given by the relation
\begin{equation}
\label{Eq: DipoleQP}
(d_{ij}^{qp})^2 = |\langle i|\sin\frac{\phi-\phi_2}{2}|j\rangle|^2 +  |\langle i|\sin\frac{\phi-\phi_2-\phi_1}{2}|j\rangle|^2
\end{equation}
Note that for a fixed frequency this mechanism predicts the same scaling of $T_1$ with the updated transition dipole as in the case of a linear loss considered above.
\section{Experimental methods}

	\begin{figure}
	\centering
	\includegraphics[width=\linewidth]{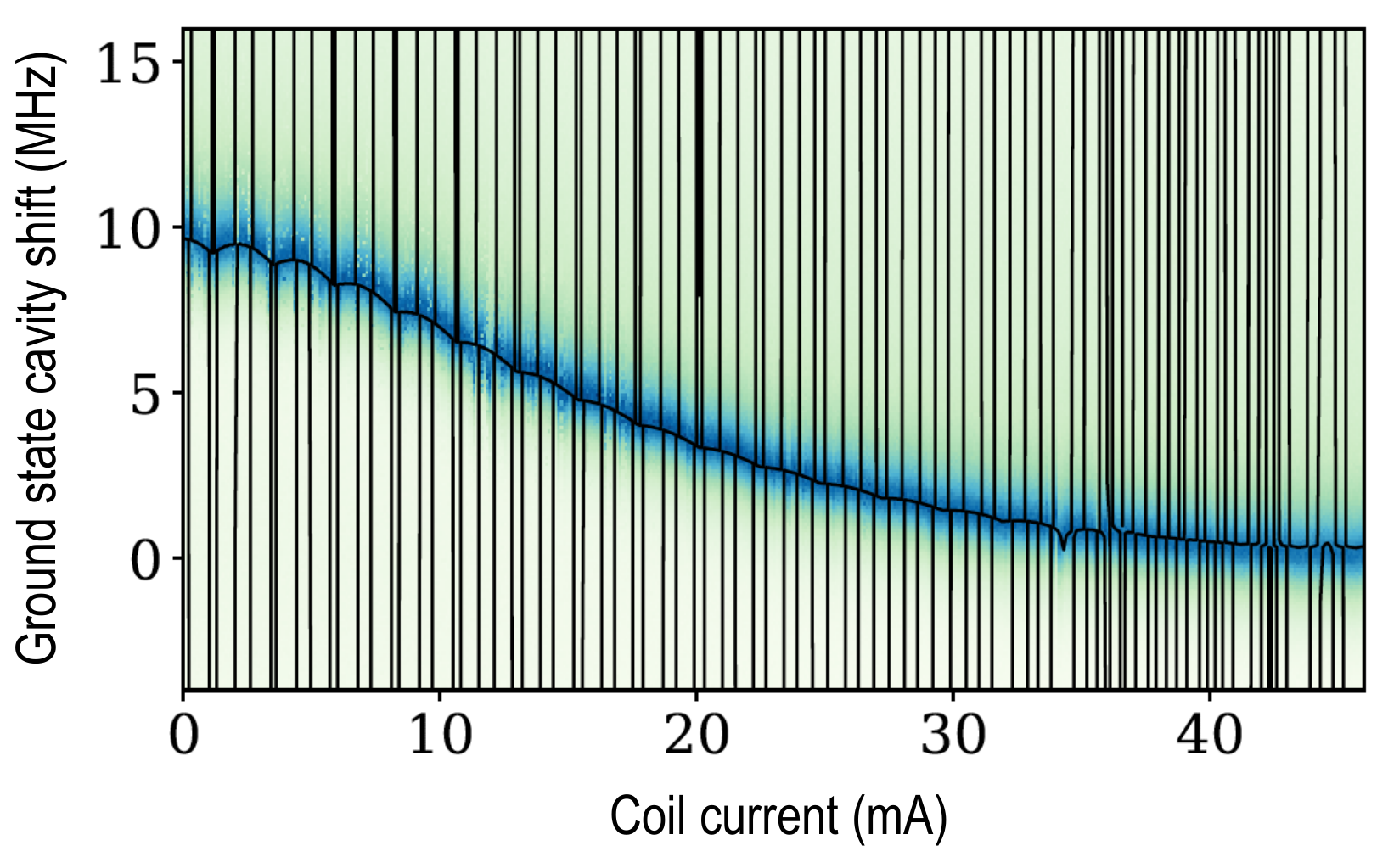}
		\caption{One-tone spectroscopy as a function of coil current. No preparation pulse was applied to the qubit, which would ideally remain in its ground state. Theory line corresponds to theory with two fit parameters: bare cavity frequency $\omega_c$ and the coupling constant $g$. The vertical black lines correspond to the fluxon transitions which cannot be seen at this large flux and frequency scale.}
		
		\label{fig:Fig6}
	\end{figure}

The fluxonium device was fabricated on an untreated Si chip by e-beam lithography using a bi-layer MMA/PMMA resist and a double-angle electron-beam deposition/oxidation/deposition sequence. The chip was mounted inside a 3D copper cavity with a single coupling port with an external quality factor of about $2,000$. The port was probed by a microwave transmission in a hanger geometry with the rest of the microwave setup being similar to that used in Ref.~\cite{manucharyan2012evidence}. The table below summarizes circuit model parameters extracted from spectroscopy data.

 \begin{table}[h]\label{Table-coupled-LC}\label{Table1}
 	\begin{tabular}{|c|c|c|c|c|c|}
 		\hline  
 		$E_{J_{\Sigma}}$ & $d$ & $E_C$ & $E_L$ &$\omega_c/2\pi$ &$g$\\
 		\hline 
 		$17.61$ & $0.13$ &	$0.55$  & $0.72$& $10.304$&$0.082$\\
 		\hline 
 	\end{tabular} 
 	\caption{Circuit parameters extracted from spectroscopy data. The atom's parameters $E_{J_{\Sigma}}, d, E_C, E_L$ were extracted by fitting the data in Fig.~\ref{fig:Fig2}a. The cavity frequency $\omega_c$ and the coupling constant $g$, defined by Eq.~\ref{cQED}, are extracted from fitting the ground state cavity shift shown in Fig.~\ref{fig:Fig6}. All energies are given in units of $\textrm{GHz}$.}
 \end{table}

\bibliography{SuperconductingCircuits.bib}

\end{document}